\documentclass[utf8]{frontiersFPHY} 

\usepackage{url,hyperref,lineno,microtype,subcaption}
\usepackage[onehalfspacing]{setspace}
\usepackage{cleveref}

\def\keyFont{\fontsize{8}{11}\helveticabold }
\def\firstAuthorLast{Yu {et~al.}} 
\def\Authors{Rong Yu\,$^{1,*}$, Haoyu Hu\,$^{2}$, Emilian M. Nica\,$^{3}$, Jian-Xin Zhu\,$^{4}$ and Qimiao Si\,$^{2}$}
\def\Address{$^{1}$Department of Physics and Beijing Key Laboratory of Opto-electronic Functional Materials and Micro-nano Devices, Renmin University of China, Beijing 100872, China \\
$^{2}$Department of Physics \& Astronomy, Center for Quantum Materials, Rice University, Houston, Texas 77005, USA \\
$^{3}$Department of Physics, Arizona State University, Box 871504, Tempe, Arizona 85287-1504, USA \\
$^{4}$Theoretical Division and Center for Integrated Nanotechnologies, Los Alamos National Laboratory, Los Alamos, New Mexico 87545, USA }
\def\corrAuthor{Rong Yu}

\def\corrEmail{rong.yu@ruc.edu.cn}

\usepackage{color,soul}

\begin{document}
\onecolumn
\firstpage{1}

\title[Orbital selectivity in iron-based superconductors]{Orbital selectivity in electron correlations and superconducting pairing of iron-based superconductors}

\author[\firstAuthorLast ]{\Authors} 
\address{} 
\correspondance{} 

\extraAuth{}

\maketitle

\begin{abstract}
Electron correlations play a central role in iron-based superconductors. In these systems,
multiple Fe $3d$-orbitals are active in the low-energy physics, and they are not all degenerate.
 For these reasons,  the role of orbital-selective correlations has been an active topic in the study of the iron-based systems.
 In this paper, we survey the recent developments on the subject.
For the normal state, we emphasize the orbital-selective Mott physics that has been extensively studied,
 especially in the iron chalcogenides, in the case of  electron filling $n \sim 6$.
 In addition, the interplay between orbital selectivity and electronic nematicity is addressed.
 For the superconducting state, we summarize the initial ideas for orbital-selective pairing, and
 discuss the recent explosive activities along this direction.
 We close with some perspectives on several emerging topics.
 These include  the evolution of the
orbital-selective correlations, magnetic and nematic orders and superconductivity
as  the electron filling factor is reduced from $6$ to $5$,
as well as
 the interplay between electron correlations
and topological bandstructure in iron-based superconductors.

\tiny
 \keyFont{ \section{Keywords:} iron-based superconductors, iron selenides, electron correlations, orbital selectivity, orbital-selective pairing}
\end{abstract}

\section{Introduction}\label{Sec:Intro}

Since the discovery of superconductivity in F-doped LaFeAsO~\citep{Hosono_JACS:2008}, the study of
 iron-based superconductors (FeSCs) has been one of the most active fields in condensed matter physics.
 The FeSCs feature a large family of materials, which are divided into two major classes,
 the iron pnictides and iron chalcogenides. The highest superconducting transition temperature ($T_c$)  is at $56$ K
 for the iron pnictides~\citep{Ren_CPL:2008,Wang_EPL:2008}, and  $65$ K or even higher in the single-layer
 iron chalcogenide~\citep{Xue_CPL:2012,He_NM:2013,Shen_Nat:2014,Wang_SB:2015,Jia_NM:2015}.
It is believed that the high-temperature superconductivity in the FeSCs originates from
electron-electron interactions
~\citep{Johnston_AP:2010,Wang_Sci:2011,Dagotto_RMP:2013,Dai_RMP:2015,Si_NRM:2016,Hirschfeld_CRP:2016}.
This inspires the consideration of similarities and differences between the FeSCs and other correlated superconductors,
especially
the cuprates~\citep{Bednorz-Mueller_ZPB:1986}.
Similar to the cuprates,
 most parent compounds of FeSCs have an antiferromagnetically (AFM)
 ordered ground state~\citep{Cruz_Nat:2008,Dai_RMP:2015},
 and the
 superconductivity emerges within a certain range of chemical doping.
 In contrast with the cuprates, most though not all the
 parent iron pnictides and iron chalcogenides are metals
 ~\citep{Johnston_AP:2010,Wang_Sci:2011,Dagotto_RMP:2013,Dai_RMP:2015,Si_NRM:2016,Hirschfeld_CRP:2016},
 typically
 exhibiting electron and hole Fermi pockets as revealed by angle resolved photoemission spectroscopy (ARPES) measurements~\citep{Yi_PRB:2009}.

 \subsection{Electron correlations in the FeSCs}
\label{Sec:electron_correlation}

There are two important issues regarding the electron correlations in FeSCs. The first concerns
their
overall strength of
electron correlations.
The parent FeSCs are bad metals~\citep{Si_PRL:2008,Abrahams_JPCM:2011},
with the room-temperature electrical resistivities exceeding
 the Mott-Ioffe-Regel limit and corresponding to  the product of Fermi wave vector and mean-free path
 being of
 order
 unity.
 The Mott-Ioffe-Regel criterion \cite{Hussey2004} signifies a system with a
 metallic ground state
 and
 with strong electron correlations.
 Further evidence for the bad metal behavior come from the optical conductivity measurement,
 which showed that the  Drude weight
is considerably reduced by the electron correlations~\citep{Qazilbash_NP:2009,HuOptics_PRL:2008,YangOptics_PRL:2009,Boris_PRL:2009,Degiorgi_NJP:2011}.
Relatedly, the effective mass of the single-electron excitations is much enhanced from their non-interacting counterpart,
with the enhancement factor ranging from $3$ to $20$
across the FeSC families~\citep{Yi_PRB:2009,Tamai_PRL:2010,Yi_NJP:2012,Liu_PRL:2013,Yi_PRL:2013,Yi_NC:2015}.
These bad-metal characteristics,
together with the existence of a large spin spectral weight observed by neutron scattering experiments
(already for the parent iron pnictides~\citep{Liu_PRB:2020,Liu_NP:2012})
and a number of other characteristics from measurements such as the X-ray emission~\citep{Gretarsson_PRB:2011}
and Raman scattering~\citep{Baum_CP:2019}
 spectroscopies,
imply that the parent FeSCs possess a considerable degree of electron correlations.
Indeed, the integrated spin spectral weight measured from the dynamical spin susceptibility
 is at the order of $3$ $\mu_B^2$ per Fe in the parent iron pnictides,
which is too large to be generated by particle-hole excitations in the Fermi-surface
nesting picture
but is consistent with the spin degrees of freedom
being dominated
by the contributions from
 the incoherent
  electronic excitations \citep{Liu_NP:2012}.
The total spin spectral weight is even larger in the iron chalcogenides \citep{Dagotto_RMP:2013,Dai_RMP:2015,Si_NRM:2016}.

All these experimental results suggest that the parent FeSCs are in the bad metal regime
which is close to a metal-to-Mott-insulator transition (MIT).
This regime can be  described by a $w$-expansion around the MIT
within the incipient Mott picture~\citep{Si_PRL:2008,Si_NJP:2009,Dai_PNAS:2009},
where $w$ is the overall fraction of the electron spectral weight that occupies the coherent itinerant part.
To the zeroth order in $w$,
the spin degrees of freedom appear in the form of
quasi-localized magnetic moments with frustrated exchange interactions;
this picture
anchors the description of the AFM order and the associated magnetic fluctuations.
The importance of such incoherent electronic excitations to the low-energy physics of the FeSCs
has also been emphasized
from related considerations
\citep{Laad_PRB:2009,Ishida_PRB:2010,Yin_NM:2011,deMedici_PRL:2014,Seo_PRL:2008,Chen_PRL:2009, Moreo_PRB:2009,Bascones_PRB:2012,Berg_NJP:2009,Lv_PRB:2010,Fang_PRB:2008,Xu_PRB:2008, Ma_PRB:2008,Wysocki_NP:2011,Uhrig_PRB:2009,Rodriguez2009,Lorenzana_NC:2011}.

\subsection{Orbital selective correlations and pairing}
\label{Sec:orbita_selective}

\begin{figure}[th!]
\begin{center}
\includegraphics[width=15cm]{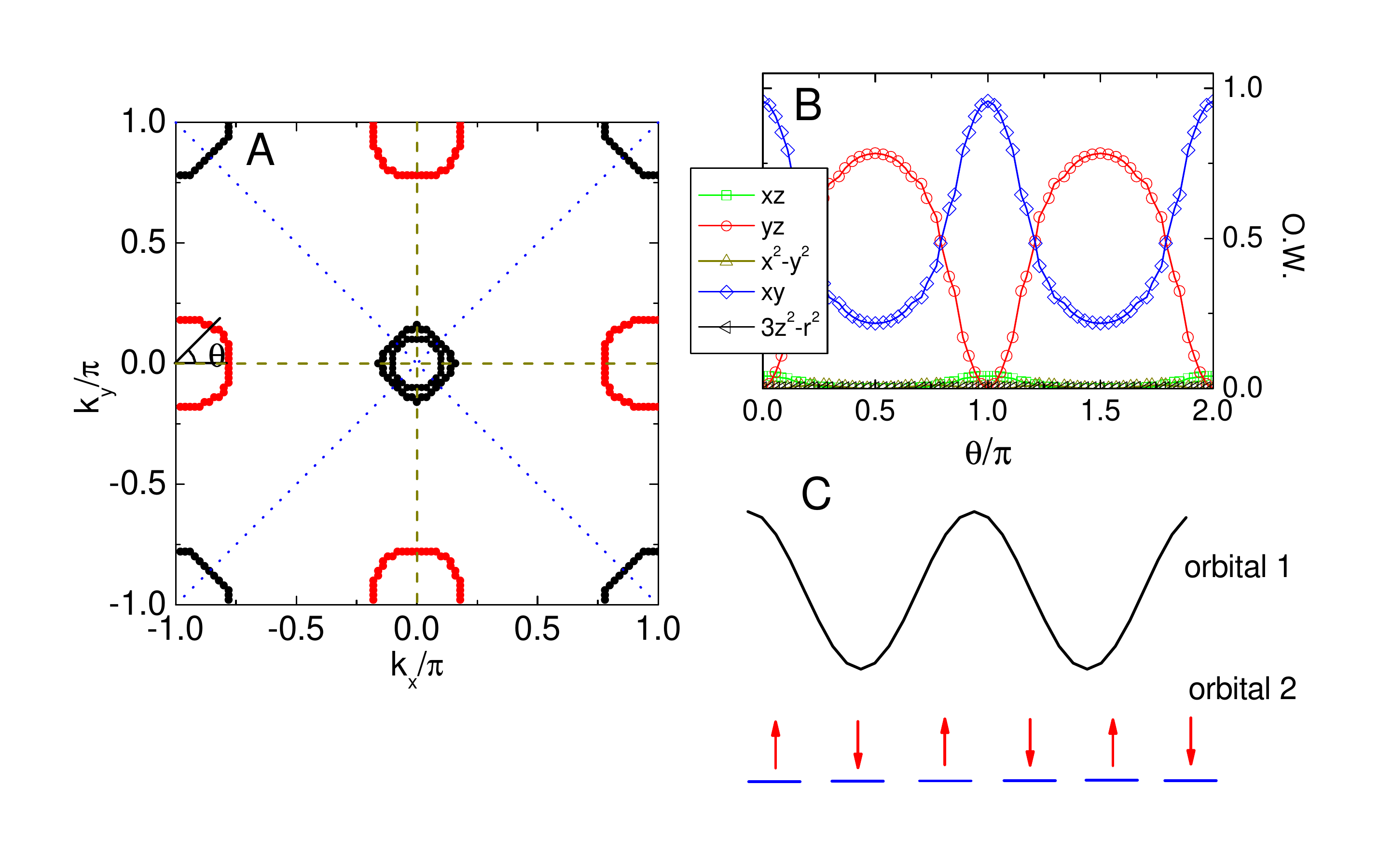}
\end{center}
\caption{(A): Fermi surface
of a five-orbital
multiorbital Hubbard
model for the
iron pnictides, consisting
of both hole (black symbols) and electron (red symbols) pockets.
The $1$-Fe Brilluion zone (BZ) is used hereafter.
(B): Orbital weights (O.W.) along the electron pocket centered at $(\pi,0)$. $\theta$ is defined in panel (A).
Adapted from Ref.~\citep{Yu_PRB:2014}.
(C): Sketch of the orbital-selective Mott phase in a two-orbital model. Orbital $1$ is metallic,
with its renormalized
 bandwidth being nonzero,
and orbital $2$ is a Mott insulator where the active degree of freedom is
a  magnetic moment localized at each site.
Note that interorbital coupling is in general nonzero in the kinetic part of the
Hamiltonian. The OSMP can develop only when the corresponding interorbital coupling
is renormalized to zero.
}\label{fig:1}
\end{figure}

The other, related, issue is the multiorbital nature
of  the low-energy electronic structure of FeSCs. As illustrated in Fig.~\ref{fig:1}(A) and (B),
 the Fermi surface of the parent FeSCs consists of several pockets, and each pocket contains
 contributions from multiple Fe $3d$ orbitals.
 The Fe ion has a valence of $+2$, corresponding to $n=6$ electrons occupying
its
  five 3$d$ orbitals.
 These orbitals are not all degenerate, and there are an even number of electrons per Fe site.
 Thus,  the MIT  in such systems, if it does take place, is expected to be
 quite distinct.
The  common tetragonal structure of FeSCs only preserves the degeneracy between the $d_{xz}$ and $d_{yz}$ orbitals.
The partially lifted orbital degeneracy may
cause the effects of
the electron correlations to be orbital dependent.
As a simple example, consider a system with two non-degenerate orbitals.
The
 bandwidths or the electron fillings in these two orbitals are
 generically
 different.
 Thus, even
 for the same Coulomb repulsion,
 the degree of electron correlations is expected to be different, and this difference
  is denoted as orbital selectivity.
  The case of extreme distinction
   corresponds to an orbital-selective Mott phase (OSMP):
As sketched in Fig.~\ref{fig:1}(C), orbital $2$ becomes a MI where the electrons are fully localized,
while orbital $1$  remains
metallic. The notion that some orbital can be driven
through
a delocalization-localization transition
while the others remain delocalized can be
traced back to the physics of Kondo destruction in $f$-electron physics~\citep{Si_2001,Coleman_JPCM:2001,Pepin_PRB:2008}.
For $d$-electron systems, the OSMP was first
considered
for Ca$_{2-x}$Sr$_{x}$RuO$_{4}$~\citep{Anisimov_EPJB:2002}
within a multiorbital model whose kinetic part is diagonal in the orbital basis, for which the orbital and band bases are the same.

An important characteristics of the FeSCs is that different orbitals are coupled to each other,
as dictated by the crystalline symmetry,
and this makes the consideration of the OSMP especially nontrivial.
Here, the
 treatment of the orbital-selective correlation effect in multiorbital models with such interorbital coupling
was introduced in Ref.~\citep{Yu_PRB:2010}.
The analysis
of Ref.~\citep{Yu_PRB:2010}
set the stage for the realization of an OSMP in the multiorbital Hubbard
models for the iron chalcogenides \citep{Yu_PRL:2013}, in which the bare Hamiltonian contains a kinetic
coupling between the $d_{xy}$ and other $3d$ orbitals of Fe. The theoretical work went together with the experimental
observation of an OSMP in several iron chalcogenides~\citep{Yi_PRL:2013,Yi_NC:2015}.
The mechanism for the suppression of interorbital coupling
by the correlation effects, which allows for the OSMP, is further clarified in terms of a Landau free-energy
 functional
in Ref.~\citep{Yu_PRB:2017}.
(Related microscopic studies have been carried out in Ref.~\citep{Komijani_PRB:2017}.)

The
recognition
of the orbital-selective correlations has led to the initial work on the
orbital-selective pairing \citep{Yu_PRB:2014}. This notion was motivated by
-- and applied to the analysis of -- the properties of the superconducting state in the under- electron-doped
NaFeAs~\citep{Zhang_PRL:2013}. In other theoretical approaches, various forms of
orbital-selective pairing were considered in the contexts of the FeSCs \citep{Yin_NP:2014, Ong_PNAS:2016}.

\subsection{Perspective and objective}
\label{Sec:remarks}

Since most of  the parent compounds are not Mott insulators (MIs), assessing the strength of electron correlations
has been an important topic since the beginning of the FeSC field.
In principle, the AFM ground state and the superconducting state nearby may
originate  from  the Fermi surface nesting mechanism of a weak-coupling theory~\citep{Dong_EPL:2008,Kuroki_PRL:2008,Graser_NJP:2009,Wang_PRL:2009,Knolle_PRB:2011}.

As outlined above, the correlation strength of the FeSCs is intermediate: Here, the Coulomb repulsion and the bandwidth
are similar in magnitude, and the competition between the electrons' itineracy and localization is the most fierce.
Spectroscopy measurements have provided ample evidence that, for the parent compounds of the FeSCs,
the incoherent part of the electron spectral weight $(1-w)$ is larger than the coherent counterpart $w$,
which provides a microscopic definition of a bad metal.
The full force of the electron correlations in the FeSCs has now become quite apparent
\cite{Si_NRM:2016,Yi_NPJQM:2017,Birgeneau2015,Wang_Zhao_PRL:2016,MYi-S-doping2015,Niu_Feng_PRB:2016,Song_NC:2016,Iimura_Hosono_PRB:2016,WatsonValenti_PRB:2016,Buchner2016,Gerber_Science:2017,Lafuerza_deMedici_prb:2017,HMiao_prb:2018,Song_prl:2019,Song_prl:2019b,YHu_XJZhou_insulating2019,Moreo_OSMP2019,Ruiz2019,Hiraishi_Hosono_2020}.
In leading towards this understanding,
the orbital selective aspects of the correlations and pairing have played an important role.
We also note that the orbital-selective correlation effects, similar to what we summarize here,
have more recently been incorporated into weak-coupling
approaches \cite{Kreisel_PRB:2017,Benfatto_NPJQM:2018}.

Recognizing that the study of the orbital-selective correlations and pairing has had explosive developments in recent years,
here we survey
the recent theoretical progress on the orbital selectivity
for both the normal and superconducting states in multiorbital models for FeSCs.
We focus on the MIT at $n=6$ and show how the Hund's coupling stabilizes a bad metal phase with a large orbital selectivity.
 Especially for the iron chalcogenides, an OSMP -- with the $d_{xy}$ orbital being Mott localized and the other $3d$ orbitals remaining itinerant --  appears in the phase diagram.
We further discuss the experimental evidences for the orbital selectivity
 as well as  the implications of the orbital-selective correlations for the
 magnetism, electronic nematicity, and superconductivity of FeSCs.
 For the superconducting state, we
 summarize how the orbital-selective superconducting pairing
 not only accounts for the strikingly large
 superconducting gap anisotropy,
 but also gives rise to novel pairing states.
 The latter
 clarifies a number of puzzles in alkaline iron selenides.

The rest of the manuscript is organized as follows: In Sec.~\ref{Sec:NormalState}
we first briefly introduce the multiorbital Hubbard model for FeSCs and outline the $U(1)$ slave-spin theory
~\cite{Yu_PRB:2012,Yu_PRB:2017}
used for studying the MIT in this model.
This approach accounts for the proper symmetry of the involved phases.
We also consider a Landau free-energy functional that demonstrates how an OSMP can be robust in spite of a nonzero bare inter-orbital kinetic hybridization.
Then we study the bad metal behavior and the OSMP that is identified
in the phase diagram of this model,
and discuss the implication for the nematici phase of iron selenide.
In Sec.~\ref{Sec:SCPairing} we set up the multiorbital $t$-$J$ model for studying the superconductivity of FeSCs,
and discuss the main results of the orbital-selective superconducting pairing and its implications.
In Sec.~\ref{Sec:Summary} we present a brief summary and an outlook for several
emerging directions.

\section{Orbital-selective correlations in the normal state of iron-based superconductors}\label{Sec:NormalState}

To study the effects of orbital-selective correlation, we consider a multiorbital Hubbard model for the FeSCs. The Hamiltonian reads
\begin{equation}
 \label{Eq:Ham_tot}
 H=H_{\rm{TB}} + H_{\mathrm{int}}.
\end{equation}
Here $H_{\rm{TB}}$ is a tight-binding model that contains multiple Fe $3d$ orbitals and
preserves the tetragonal lattice symmetry of the FeSCs in the normal state.
The tight-binding parameters are obtained by fitting to the DFT bandstructure of specific compounds.
A number of tight-binding models, ranging from two- to five-orbital models have been proposed for FeSCs~\citep{Raghu_PRB:2008,Kuroki_PRL:2008,Yu_PRB:2009,Graser_NJP:2009,Daghofer_PRB:2010,Graser_PRB:2010}.
In principle, any of these models can be used to illustrate the correlation effects.
In practice, we adopt the more realistic five-orbital models, which
capture
 the salient features of the electronic structure and Fermi surface
and facilitate
 a
direct comparison to the experimental results.
As already stressed, the tetragonal symmetry dictates that interorbital hopping amplitudes are allowed and the fitted tight-binding
parameters do show that such hopping amplitudes are nonzero. Specifically, the non-interacting part of the Hamiltonian is written as
\begin{equation}
 \label{Eq:Ham_0} H_{\rm{TB}}=\frac{1}{2}\sum_{ij\alpha\beta\sigma} t^{\alpha\beta}_{ij}
 d^\dagger_{i\alpha\sigma} d_{j\beta\sigma} + \sum_{i\alpha\sigma} (\Lambda_\alpha-\mu) d^\dagger_{i\alpha\sigma} d_{i\alpha\sigma} \, .
\end{equation}
Here,
$d^\dagger_{i\alpha\sigma}$ creates an electron in orbital $\alpha$ ($=1,...,5$)
with spin $\sigma$ at site $i$; $t^{\alpha\beta}_{ij}$, with $i \ne j$,
are the tight-binding parameters, with those for $\alpha \ne \beta$ describing the inter-orbital couplings;
and $\Lambda_\alpha$
refers to the orbital-dependent energy levels,
 associated with the crystal field
 splittings,
  and is diagonal in the orbital basis.
 In particular,
 the C$_4$ symmetry dictates that $d_{xz}$ and $d_{yz}$ are degenerate, but no symmetry enforces any degeneracy between
 the $d_{xy}$ orbital
 and the other orbitals. Indeed,
 for any orbital $\beta \ne xy$, $\Lambda_{xy,\beta} \equiv \Lambda_{xy} - \Lambda_{\beta} \ne 0$.
The chemical potential $\mu$  controls the total electron number $n$
that occupies the $3d$ orbitals of each Fe site. In the model,
$n=6$ for the parent (undoped) compound.

The onsite interaction $H_{\rm{int}}$ takes the following form,
\begin{eqnarray}
 \label{Eq:Ham_int} H_{\rm{int}} &=  \frac{U}{2} \sum_{i,\alpha,\sigma}n_{i\alpha\sigma}n_{i\alpha\bar{\sigma}} &+
 \sum_{i,\alpha<\beta,\sigma} \left\{ U^\prime n_{i\alpha\sigma} n_{i\beta\bar{\sigma}}
 + (U^\prime-J_{\rm{H}}) n_{i\alpha\sigma} n_{i\beta\sigma} \right\} \nonumber\\
& &
-J_{\rm{H}}\sum_{i,\alpha<\beta,\sigma}\left(d^\dagger_{i\alpha\sigma}d_{i\alpha\bar{\sigma}} d^\dagger_{i\beta\bar{\sigma}}d_{i\beta\sigma}
 +d^\dagger_{i\alpha\sigma}d^\dagger_{i\alpha\bar{\sigma}}
 d_{i\beta\sigma}d_{i\beta\bar{\sigma}} \right),
\end{eqnarray}
where $n_{i\alpha\sigma}=d^\dagger_{i\alpha\sigma} d_{i\alpha\sigma}$.
Here,
$U$
and
$U^\prime$
denote the intraorbital
and interorbital repulsion,
respectively,
and
$J_{\rm{H}}$  is
the Hund's rule exchange coupling.
These coupling parameters satisfy
$U^\prime=U-2J_{\rm{H}}$~\citep{Castellani_PRB:1978}.

\subsection{The $U(1)$ slave-spin theory}
\label{Sec:NormalState_U1}

The multiorbital system described by the model in Eq.~\eqref{Eq:Ham_tot} undergoes a MIT driven
by the electron correlations at any commensurate electron filling. This transition can be
studied by using a $U(1)$ slave-spin theory~\citep{Yu_PRB:2012,Yu_PRB:2017}.
In this subsection we summarize the theoretical approach for the MIT and show
that, besides the conventional metallic and Mott insulating phases,
there is
 also an
OSMP as the ground state of the system.

Slave-particle (or parton) construction has a long history in the study of correlated systems
 \citep{Barnes_JPF:1976, Coleman_PRB:1984, Read_JPC:1983, Kotliar_PRL:1986}.
 More recent variations include slave rotor \citep{Florens_PRB:2004} and  $Z_2$ slave-spin
\citep{deMedici_PRB:2005} formulations. For the purpose of describing the MIT,
the $Z_2$ gauge structure~\citep{deMedici_PRB:2005,deMedici_PRB:2010,Ruegg_PRB:2010}
is problematic given that the MIT concerns the (de)localization of a $U(1)$-symmetric
 charge degrees of freedom~\cite{Nandkishore_PRB:2012}.
 The $U(1)$ slave-spin formulation is more suitable,
given that it captures the symmetry of the pertinent phases.

In the $U(1)$ slave-spin formulation~\citep{Yu_PRB:2012,Yu_PRB:2017},
the electron creation operator is represented as
\begin{equation}
 \label{Eq:SSCreate} d^\dagger_{i\alpha\sigma} = S^+_{i\alpha\sigma} f^\dagger_{i\alpha\sigma}.
\end{equation}
Here the XY component of a quantum $S=1/2$ spin operator ($S^+_{i\alpha\sigma}$) is used to represent
the charge degree of freedom of the electron at each site $i$, for each orbital $\alpha$
and each spin flavor $\sigma$. Correspondingly, the fermionic ``spinon'' operator,
$f^\dagger_{i\alpha\sigma}$, is used to carry the spin degree of freedom.
To restrict the Hilbert space to the physical one, a local constraint
\begin{equation}
 \label{Eq:constraint} S^z_{i\alpha\sigma} = f^\dagger_{i\alpha\sigma} f_{i\alpha\sigma} - \frac{1}{2}
\end{equation}
is implemented. This representation contains a $U(1)$ gauge redundancy corresponding to
$f^\dagger_{i\alpha\sigma}\rightarrow f^\dagger_{i\alpha\sigma} e^{-i\theta_{i\alpha\sigma}}$
and $S^+_{i\alpha\sigma}\rightarrow S^+_{i\alpha\sigma} e^{i\theta_{i\alpha\sigma}}$.
In parallel
to the slave-rotor approach~\citep{Florens_PRB:2004}, in this representation,
the slave spins carry the $U(1)$ charge.
Correspondingly,
the $U(1)$ slave-spin theory
can naturally describe the MIT,
including
 in multiorbital settings.

To construct a saddle-point theory one has to work within the Schwinger boson representation of the slave spins. A detailed derivation of the saddle-point equations can be found in Refs.~\citep{Yu_PRB:2012,Yu_PRB:2017}. Here, for conciseness, we will mostly
stay in the slave-spin representation and simply describe the main results.
To ensure that the quasiparticle spectral weight in the non-interacting
limit is normalized to $1$ at the
saddle point level,
and in analogy to
the standard treatment in the slave-boson theory~\citep{Kotliar_PRL:1986},
we
 define a dressed operator:
\begin{equation}
 \label{Eq:Zdagger} \hat{z}^\dagger_{i\alpha\sigma} = P^+_{i\alpha\sigma} S^+_{i\alpha\sigma}
 P^-_{i\alpha\sigma},
\end{equation}
where the projectors $P^\pm_{i\alpha\sigma}=1/\sqrt{1/2+\delta \pm S^z_{i\alpha\sigma} }$, and $\delta$ is an infinitesimal positive
number to regulate $P^\pm_{i\alpha\sigma}$.
Next we rewrite Eq.~\eqref{Eq:SSCreate} with the dressed operator to
$d^\dagger_{i\alpha\sigma}=\hat{z}^\dagger_{i\alpha\sigma} f^\dagger_{i\alpha\sigma}$.
The Hamiltonian, Eq.~ \eqref{Eq:Ham_tot},
is then effectively rewritten as
\begin{eqnarray}
\label{Eq:HamSS} H &=& \frac{1}{2}\sum_{ij\alpha\beta\sigma} t^{\alpha\beta}_{ij}
 \hat{z}^\dagger_{i\alpha\sigma} \hat{z}_{j\beta\sigma} f^\dagger_{i\alpha\sigma} f_{j\beta\sigma}
 + \sum_{i\alpha\sigma}  (\Lambda_\alpha -\mu) f^\dagger_{i\alpha\sigma}
 f_{i\alpha\sigma}
 \nonumber\\
 &&
- \lambda_{i\alpha\sigma}[f^\dagger_{i\alpha\sigma}
 f_{i\alpha\sigma}-
 S^z_{i\alpha\sigma}] + H^S_{\mathrm{int}}.
\end{eqnarray}
Here, we have introduced the Lagrange multiplier $\lambda_{i\alpha\sigma}$ to
enforce the constraint in Eq.~\eqref{Eq:constraint}.
In addition,  $H^S_{\mathrm{int}}$ is the interaction Hamiltonian, Eq.~\eqref{Eq:Ham_int},
 rewritten
 in terms of
the slave-spin operators~\citep{Yu_PRB:2010}
as follows
\begin{eqnarray}\label{e.hsint}
H^S_{int} &=& \sum_{\mathbf{i}} \left\{\frac{U'}{2}\left(\sum_{\alpha\sigma} S^z_{\mathbf{i}\alpha\sigma}\right)^2 + \frac{U-U'}{2} \sum_{\alpha}\left(\sum_{\sigma} S^z_{\mathbf{i}\alpha\sigma}\right)^2\right. \nonumber\\ &-& \frac{J_{\rm{H}}}{2} \sum_{\sigma}\left(\sum_{\alpha} S^z_{\mathbf{i}\alpha\sigma}\right)^2 -J_{\rm{H}}
\sum_{\alpha<\beta} \left[S^+_{\mathbf{i}\alpha\uparrow} S^-_{\mathbf{i}\alpha\downarrow}
S^+_{\mathbf{i}\beta\downarrow} S^-_{\mathbf{i}\beta\uparrow} \right. \nonumber\\
&-&\left.\left.
S^+_{\mathbf{i}\alpha\uparrow} S^+_{\mathbf{i}\alpha\downarrow}
S^-_{\mathbf{i}\beta\downarrow} S^-_{\mathbf{i}\beta\uparrow} +
\rm{H.c.}\right]\right\}.
\end{eqnarray}
One practical way is to
 neglect the spin flip terms in Eq.~\eqref{e.hsint}
 without affecting the qualitative results
 ~\citep{Yu_PRB:2012}.
The quasiparticle spectral weight is defined as
\begin{equation}
\label{Eq:qpWeightZ}
Z_{i\alpha\sigma}=
|z_{i\alpha\sigma}|^2 \equiv
|\langle \hat{z}_{i\alpha\sigma}\rangle|^2 .
\end{equation}
A metallic phase corresponds to $Z_{i\alpha\sigma}>0$
for all orbitals, and a Mott insulator corresponds to $Z_{i\alpha\sigma}=0$
in all orbitals with a gapless spinon spectrum.

At the saddle-point level,  the slave-spin and spinon operators
are decomposed and the constraint is treated
 on average,
We obtain two
effective
Hamiltonians for the spinons and the slave spins, respectively:
\begin{eqnarray}
 \label{Eq:Hfmf}
 H^{\mathrm{eff}}_f &=&  \sum_{k\alpha\beta}\left[ \epsilon^{\alpha\beta}_{k} \langle \tilde{z}^\dagger_\alpha \rangle
  \langle \tilde{z}_\beta \rangle + \delta_{\alpha\beta}(\Lambda_\alpha-\lambda_\alpha+\tilde{\mu}_\alpha-\mu)\right] f^\dagger_{k\alpha} f_{k\beta},\\
 \label{Eq:HSSmf}
 H^{\mathrm{eff}}_{S} &=& \sum_{\alpha\beta} \left[Q^f_{\alpha\beta}
 \left(\langle \tilde{z}^\dagger_\alpha\rangle \tilde{z}_\beta+ \langle \tilde{z}_\beta\rangle \tilde{z}^\dagger_\alpha\right)
 + \delta_{\alpha\beta}\lambda_\alpha S^z_{\alpha}\right] + H^S_{\mathrm{int}},
\end{eqnarray}
where $\delta_{\alpha\beta}$ is Kronecker's delta function,
$\epsilon^{\alpha\beta}_{k}=\frac{1}{N}\sum_{ij\sigma} t^{\alpha\beta}_{ij} e^{ik(r_i-r_j)}$,
$Q^f_{\alpha\beta}
=
 \sum_{k\sigma}\epsilon^{\alpha\beta}_k\langle f^\dagger_{k\alpha\sigma}
f_{k\beta\sigma}\rangle/2$, and $\tilde{z}^\dagger_\alpha
=
 \langle P^+_\alpha\rangle S^+_\alpha \langle P^-_\alpha\rangle$. Finally,
$\tilde{\mu}_\alpha$ is an effective onsite potential defined as
$\tilde{\mu}_\alpha = 2\bar{\epsilon}_\alpha \eta_\alpha$,
where
$\bar{\epsilon}_\alpha = \sum_\beta\left( Q^f_{\alpha\beta} \langle\tilde{z}_\alpha^\dagger\rangle \langle \tilde{z}_\beta \rangle  + \rm{c.c.} \right)$
and
$\eta_\alpha = (2n^f_\alpha-1)/[4n^f_\alpha(1-n^f_\alpha)]$,
with $n^f_\alpha=\frac{1}{N}\sum_k \langle f^\dagger_{k\alpha} f_{k\alpha} \rangle$.

We study the MIT in the paramagnetic phase preserving the translational symmetry,
and can hence drop the spin and/or site indices of the slave spins and the Lagrange multiplier
in the
saddle-point equations, Eqs.~\eqref{Eq:Hfmf} and \eqref{Eq:HSSmf}. The
parameters $z_\alpha$ and $\lambda_\alpha$ are then solved self-consistently.

\subsection{Landau free-energy functional for orbital-selective Mott physics}
\label{Sec:NormalState_Landau}

As described earlier
and illustrated in
Fig.~\ref{fig:1}(C),
the OSMP can develop only when (at least) one of the orbitals becomes localized, while the others remain delocalized.
How can this be possible in a multiorbital model with nonzero bare interorbital coupling between
orbitals?
While the $U(1)$ slave-spin approach found an affirmative answer, it is important to ask whether
the result of this microscopic approach is robust. To do so,
we construct a Landau theory based on the slave-spin formulation~\citep{Yu_PRB:2017}.
We start from the saddle-point Hamiltonians Eqs.~\eqref{Eq:Hfmf} and \eqref{Eq:HSSmf}.
Consider first Eq.~\eqref{Eq:Hfmf}, where the kinetic hybridization between two different orbitals
$\alpha \ne \beta$ is $W_{k}^{\alpha \beta} f^\dagger_{k\alpha}f_{k\beta} $,
with $W_{k}^{\alpha \beta}= \epsilon^{\alpha\beta}_{k} \langle \tilde{z}^\dagger_\alpha \rangle
\langle \tilde{z}_\beta \rangle
\propto \langle \tilde{z}^\dagger_\alpha \rangle \langle \tilde{z}_\beta \rangle$.
Because $f^\dagger_{k\alpha}f_{k\beta}$ is conjugate to $W_{k}^{\alpha \beta}$, from the linear response theory, it is easy to show that $\langle f^\dagger_{k\alpha}f_{k\beta}\rangle \propto
W_k^{\alpha \beta} \propto \langle \tilde{z}_\alpha \rangle \langle \tilde{z}^\dagger_\beta \rangle$.
As a result, the kinetic hybridization of the spinons is
\begin{equation}
\label{Eq:KHf}
 \langle H^{\rm{eff}}_f\rangle_{\alpha\beta}=
 \sum_k
 \epsilon^{\alpha\beta}_{k}
\langle \tilde{z}^\dagger_\alpha \rangle \langle \tilde{z}_\beta \rangle \langle
f^\dagger_{k\alpha}f_{k\beta} \rangle \propto |\langle \tilde{z}_\alpha \rangle|^2 |\langle \tilde{z}_\beta \rangle|^2.
\end{equation}
Next for Eq.~\eqref{Eq:HSSmf}, we can define an effective field of $h_{\alpha} = \sum_{\beta} Q^f_{\alpha\beta} \langle \tilde{z}_{\beta}\rangle$.
For similar reasoning as
 mentioned above, we obtain $Q^f_{\alpha\beta}\propto \langle \tilde{z}_\alpha \rangle \langle \tilde{z}^\dagger_\beta \rangle$, which leads to
\begin{equation}
\label{Eq:KHS}
\langle H^{\rm{eff}}_S \rangle_{\alpha\beta}
\rightarrow |\langle \tilde{z}_\alpha \rangle|^2 |\langle \tilde{z}_\beta \rangle|^2.
\end{equation}

\begin{figure}[h!]
\begin{center}
\includegraphics[width=12cm]{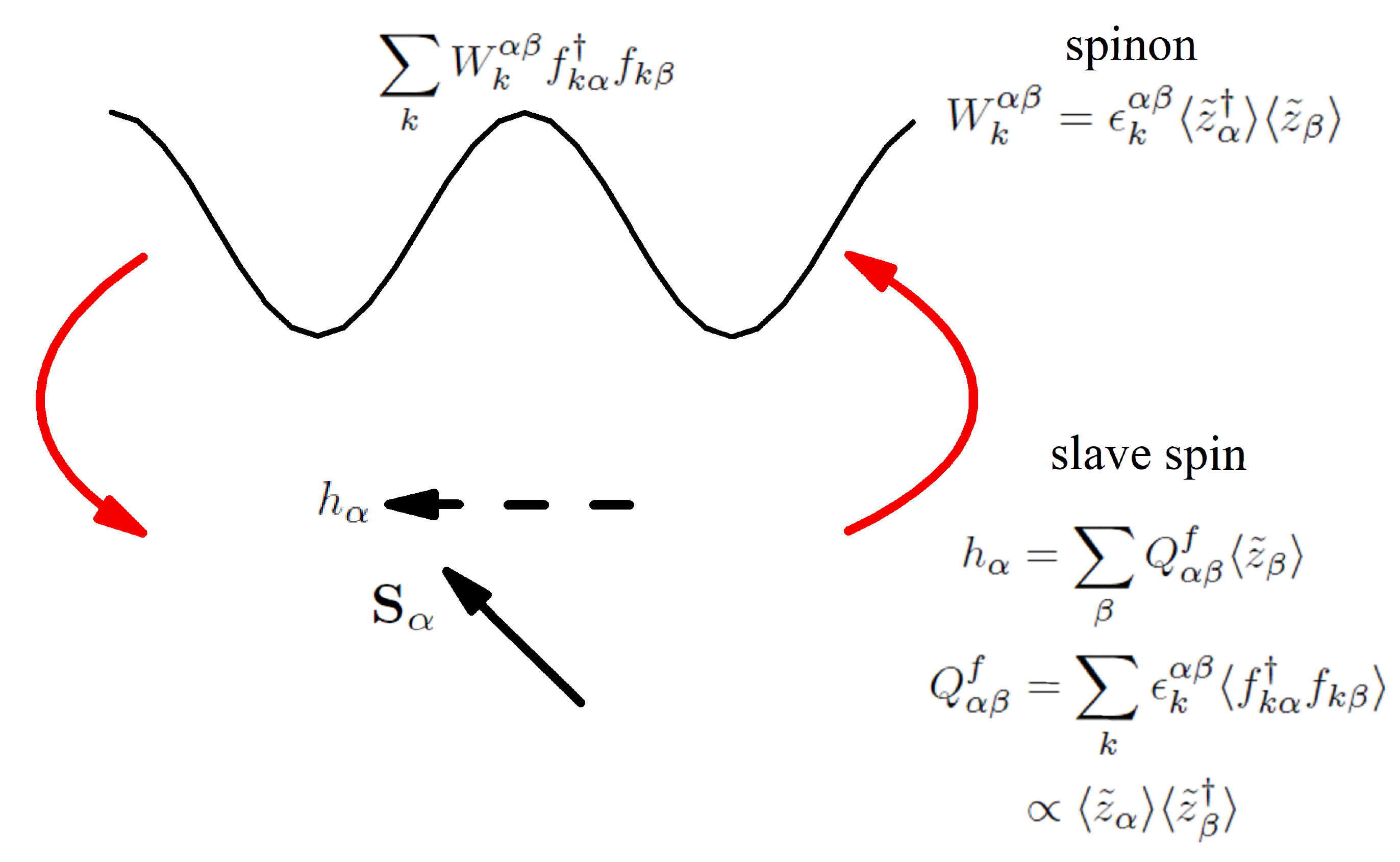}
\end{center}
\caption{Illustrating
the effect of the interorbital kinetic hybridization in the $U(1)$ slave-spin theory.
The black curve shows the effective spinon dispersion,
which is generated by $ W_{k}^{\alpha\beta}f^{\dagger}_{\alpha\sigma} f_{\beta\sigma}$.
(The physical spin index $\sigma$ is suppressed in the figure legends.)
Meanwhile, the slave-spin $\mathbf{S}_\alpha$ experiences a local field,
$h_\alpha=\sum_\beta Q^f_{\alpha\beta}\langle\tilde{z}_\beta\rangle$,
where $Q^f_{\alpha\beta}\propto\langle\tilde{z}_\alpha\rangle \langle\tilde{z}^\dagger_\beta\rangle$.
The red arrows indicate
the self-consistency between $W_{k}^{\alpha\beta}$ and $h_{\alpha}$,
which results in a biquadratic interorbital coupling as shown in Eqs.~\eqref{Eq:KHf} and ~\eqref{Eq:KHS}.
Adapted from Ref.~\citep{Yu_PRB:2017}}\label{fig:2}
\end{figure}

Note that Eqs.~\eqref{Eq:KHf} and \eqref{Eq:KHS} are natural consequence of Eqs.~\eqref{Eq:Hfmf} and \eqref{Eq:HSSmf},
and this self-consistent procedure of the saddle-point theory is illustrated in Fig.~\ref{fig:2}. Based on Eqs.~\eqref{Eq:KHf} and \eqref{Eq:KHS} we can construct a Landau free-energy functional in terms of
the quasiparticle weights, $z_\alpha$.
For simplicity of notation, we consider a two-orbital model, but our analysis straightforwardly
generalizes to the case of more than two orbitals. The free-energy density
 reads
\begin{equation}
\label{Eq:GL} f=\sum_{\alpha=1,2} \left(r_\alpha | z_\alpha |^2 + u_\alpha | z_\alpha |^4\right) + v| z_1 |^2| z_2 |^2,
\end{equation}
in which the quadratic terms $r_\alpha | z_\alpha |^2$ arise from the kinetic energy of the
saddle-point Hamiltonian in Eq.~\eqref{Eq:HSSmf} [as well as in Eq.~\eqref{Eq:Hfmf}].
The biquadratic coupling $v$ term comes from the kinetic hybridization in Eqs.~\eqref{Eq:KHf} and \eqref{Eq:KHS}.
The biquadratic nature of this intra-orbital coupling -- as opposed  to the bilinear  form --
is crucial to the stabilization of an OSMP.
This can be seen
 by
taking the derivatives
of Eq.~\eqref{Eq:GL}
with respect to $|z_\alpha|$:
Besides the conventional metallic phase with $|z_1|\neq0$, $|z_2|\neq0$ and the MI with $|z_1|=|z_2|=0$, Eq.~\eqref{Eq:GL} supports a third solution with
$|z_1|=0$, $|z_2|=\sqrt{-\frac{r_2}{2u_2}}$
 (or $|z_2|=0$, $|z_1|=\sqrt{-\frac{r_1}{2u_1}}$), which
 corresponds to an OSMP.

\subsection{Orbital-selective Mott physics in FeSCs}\label{Sec:NormalState_OSMP}

We now turn to microscopic studies of the MIT.
A realistic microscopic model for FeSCs is described in Eq.~\eqref{Eq:Ham_tot}. Owing to its multiorbital
 nature,
  the MIT in this model shows unique features. First, the parent compound corresponds to $n=6$, containing an even number of electrons per Fe ion.
 Since Mott transition is more readily defined in systems with an odd number of electrons per unit cell, this makes it
 a nontrivial question whether the model generally supports a MI in the strong correlation limit.
 Second, besides the Coulomb repulsion $U$, the Hund's rule coupling $J_{\rm{H}}$ also plays,
 a very important role in settling the ground states of the model and
 as we will see, plays an important role in realizing an OSMP.

\begin{figure}[h!]
\begin{center}
\includegraphics[width=12cm]{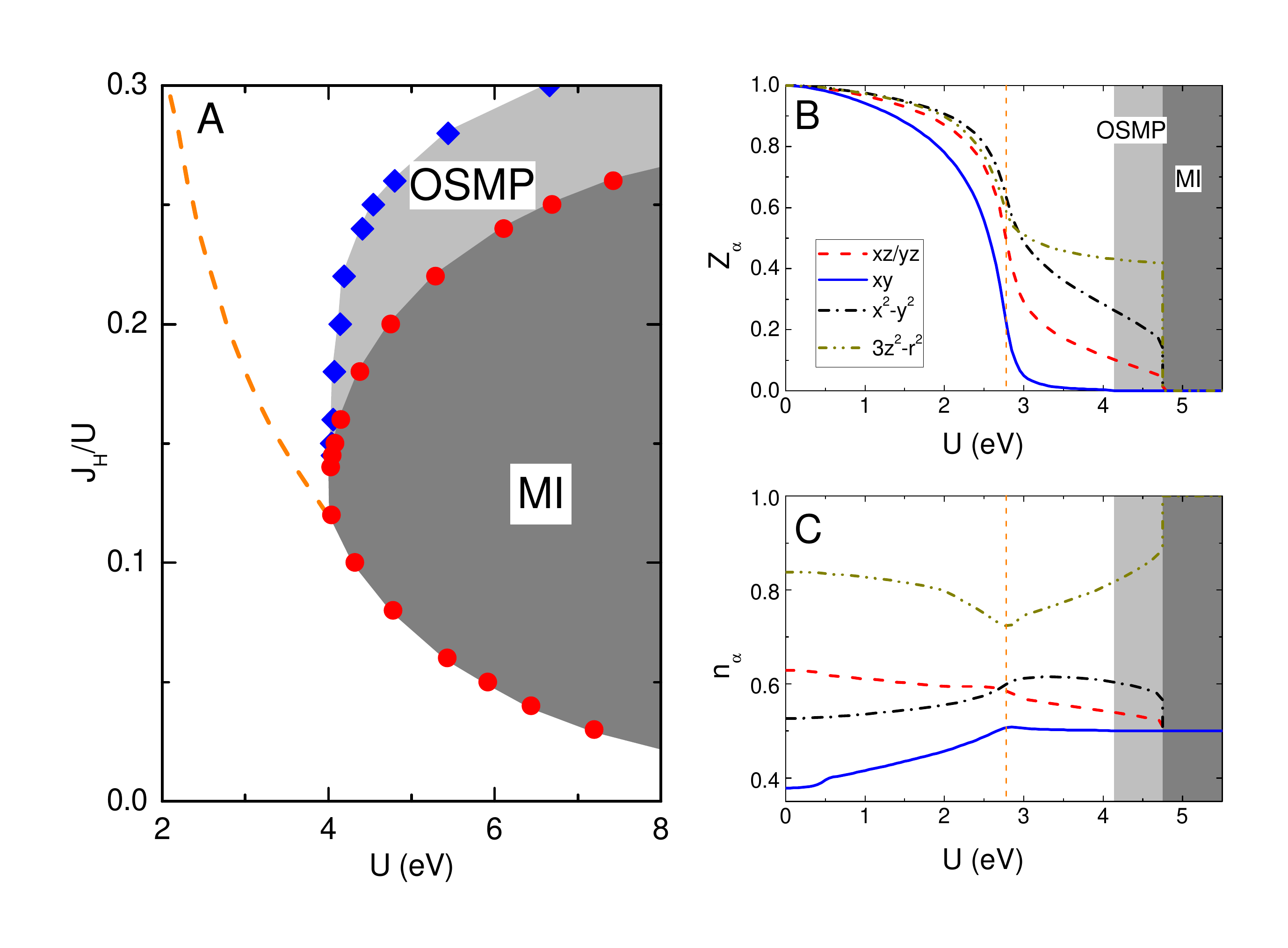}
\end{center}
\caption{(A): Ground-state phase diagram of the five-orbital
multiorbital model for alkaline iron selenides
at commensurate filling $n=6$. The dark and light grey regions correspond to the MI
and OSMP, respectively. The orange dashed line refers to a crossover
between the weakly correlated metal (WCM) and bad metal.
(B) and (C): The evolution of the orbital resolved quasiparticle spectral weights
[in (B)]
and electron filling factor per spin [in (C)]
 with $U$ in the five-orbital model at
$J_H/U=0.2$.
 Adapted from Ref.~\cite{Yu_PRL:2013}}\label{fig:3}
\end{figure}

The MIT in the multiorbital model for FeSCs at $n=6$ has been studied by using the slave-spin methods introduced in Sec.~\ref{Sec:NormalState_U1}. In the following, we present the resulting phase diagram for alkaline iron selenides.
As shown in Fig.~\ref{fig:3}(A), a MI is generally stabilized in the phase diagram when the Coulomb repulsion $U$ is sufficiently strong. The critical value $U_c$ for the Mott transition displays a non-monotonic dependence on the Hund' coupling $J_{\rm{H}}$. This is a general feature of the MIT of the multiorbital Hubbard model away from half-filling and can be understood as follows~\citep{Yu_PRB:2012}: Naively the MIT takes place when the Mott gap at the
atomic limit $G_A$ approaches
the bare
 bandwidth of the tight-binding model $D$. In the weak $J_{\rm{H}}$ limit, the MI is dominated by the low-spin $S=1$ configuration and correspondingly
 $G_A\sim
U+J_{\rm{H}}$. This leads to $U_c\sim D/(1+J_{\rm{H}}/U)$, namely, $U_c$ decreases with $J_{\rm{H}}/U$.
  On the other hand, for large $J_{\rm{H}}/U$, the high-spin $S=2$ configuration dominates
  in the MI
  state
  and $G_A\sim U-3J_{\rm{H}}$. Consequently, $U_c\sim D/(1-3J_{\rm{H}}/U)$, increases with $J_{\rm{H}}/U$.

Importantly,
 the Hund's coupling already strongly affects the properties of the metallic state. For $J_{\rm{H}}/U \gtrsim 0.1$,
 the system undergoes a crossover from a weakly-correlated metal (WCM) to a
bad metal with increasing $U$
[shown as the dashed line in Fig.~\ref{fig:3}(A)].
As illustrated in Fig.~\ref{fig:3}(B), the orbital resolved quasiparticle spectral weight $Z_{\alpha}$ in each orbital $\alpha$ rapidly drops across this crossover. Inside the
bad metal phase, $Z_{\alpha}$ becomes strongly orbital dependent. This large orbital-selective correlation is rather surprising given that the strength of the onsite interaction is identical for each orbital. To understand the strong orbital selectivity, note that the Hund's coupling suppresses interorbital correlations. For large Hund's coupling, this causes an effective orbital decoupling between any two non-degenerate orbitals and hence promotes the $S=2$ high-spin configuration in the
bad metal regime. As a result of the orbital decoupling, the correlation effect in each non-degenerate orbital depends on its filling factor $n_\alpha$. This is clearly seen in Fig.~\ref{fig:3}(B) and (C): the $d_{xy}$ orbital experiences the strongest correlation effect and $n_{xy}$ is the closest to $1/2$; while the least correlated $3z^2-r^2$ orbital has the largest filling away from $1/2$.

Further increasing $U$ in the
bad metal phase, the system undergoes a transition to
the
OSMP. In this phase the $d_{xy}$ orbital is fully Mott localized ($Z_{xy}=0$) whereas the electrons in other Fe $3d$ orbitals are still itinerant ($Z_\alpha>0$). Besides the aforementioned orbital decoupling effect, several other factors are also important for stabilizing the OSMP. First, the
bare bandwidth projected to the $d_{xy}$ orbital is smaller than
that of the  other orbitals. Second, the orbital fluctuations in the degenerate $d_{xz}$ and $d_{yz}$ orbitals
make the threshold interaction needed for their Mott localization larger than that for the non-degenerate $d_{xy}$ orbital,
although the filling factors of these three orbitals are all close to $1/2$. Taking into account all these factors, the $d_{xy}$
orbital has the lowest interaction threshold for the Mott transition, at which the the OSMP takes place.
Because only the $d_{xy}$ orbital is Mott localized, the OSMP survives
nonzero
 doping, while the MI can only
be stabilized at commensurate fillings. It is worth noting that for a particular system, whether an OSMP
is stabilized depends on the competition of above factors. For example, the OSMP
is stabilized in the model for K$_x$Fe$_{2-y}$Se$_2$ but not
that for
 LaOFeAs.\citep{Yu_PRL:2013,Yu_PRB:2012} This is mainly because the bare bandwidth of the $d_{xy}$ orbital is
 sufficiently
 smaller than that of
 the
 other orbitals in K$_x$Fe$_{2-y}$Se$_2$, but the
 difference in the bare bandwidths is less pronounced in the case of
  LaOFeAs. Even though an OSMP is not stabilized as the true ground state
  for this iron pnictide,
  it is energetically competitive~\citep{Yu_PRB:2012}. As such, the OSMP can be viewed
  as the anchoring point for the strong orbital-selective correlation effects associated with the bad metal behavior
  both in the case of the iron selenides and iron pnictides.

The OSMP
is supported by
additional theoretical studies.
 Besides the K$_x$Fe$_{2-y}$Se$_2$ system, evidences of strong orbital-selective correlations and OSMP
 have been found in several other multiorbital models for FeSCs~\citep{deMedici_PRL:2014,Rincon_PRL:2014,Backes_PRB:2015}.
 Experimentally, ARPES measurements provide clear evidence~\citep{Yi_PRL:2013,Yi_NC:2015} for OSMP in iron chalcogenides.
As temperature goes above about $100$ K, the spectral weight for the $d_{xy}$ orbital vanishes,
while that for the $d_{xz/yz}$ orbitals does not
change much~\citep{Yi_NC:2015}.
The behavior is similar
 for all the iron chalcogenides studied as well as for the alkaline iron pnictide~\citep{Miao_PRB:2016},
 which
  suggests a universal crossover to the OSMP
in FeSCs ~\citep{Yi_NC:2015}.
  Additional evidence for the OSMP
has come from THz spectroscopy~\citep{Wang_NC:2014}, Hall measurements~\citep{Ding_PRB:2014},
pump-probe
 spectroscopy~\citep{Li_PRB:2014} and high-pressure transport measurements~\citep{Gao_PRB:2014}.
Moreover, a variety of other Fe-based systems have been studied for the orbital-selective Mott behavior
 ~\citep{Birgeneau2015,Wang_Zhao_PRL:2016,MYi-S-doping2015,Niu_Feng_PRB:2016,Song_NC:2016,Iimura_Hosono_PRB:2016,Hiraishi_Hosono_2020}.
We note in passing that related orbital-selective correlation effects have recently been discussed
in the multiorbital $5f$-based actinide systems ~\citep{Chen_PRL:2019,Giannakis_SA:2019}.

\subsection{Orbital selectivity in the nematic phase of FeSe}\label{Sec:NormalState_NemOSM}

In most iron pnictides, lowering the temperature in the parent compounds gives rise to a tetragonal-to-orthorhombic structural transition at $T_s$. Right at or slightly below $T_s$, the system exhibits a transition to a collinear $(\pi,0)$ AFM state~\citep{Dai_RMP:2015}. A likely explanation for the nematicity below $T_s$ is an Ising-nematic transition of quasi-localized magnetic moments, described by an effective $J_1-J_2$-like model~\citep{Si_PRL:2008,Dai_PNAS:2009,Fang_PRB:2008,Xu_PRB:2008}.
	
Experiments in bulk FeSe do not seem to fit into this framework. Under ambient pressure,
a nematic phase without an AFM long-range order is stabilized in the bulk FeSe below the structural transition
at $T_s \approx 90$ K.  This suggests an unusual type of magnetism
 in the ground state~\citep{Yu_PRL:2015,Wang_NP:2015}. The nematic order parameter linearly couples to
 the splitting between the $d_{xz}$ and $d_{yz}$ orbitals, which can be experimentally detected. In the nematic phase of FeSe,
 ARPES measurements find this splitting to be momentum dependent,
 and the splittings at both the $\Gamma$ and M points of the Brilluion zone (BZ)~\citep{Watson_PRB:2015,Watson_PRB:2016,Zhang_PRB:2016,Liu_PRX:2018,Rhodes_Kim_FeSe_PRB:2018,Kushnirenko_Borisenko_PRB:2018,Yi_PRX:2019,Huh_CP:2020},
 $\Delta E_\Gamma$ and $\Delta E_{\rm{M}}$, respectively, are
 relatively small (less than $50$ meV).
Meanwhile, recent scanning tunneling microscopy (STM) experiments have revealed a strong orbital selectivity~\citep{Davis_Sci:2017,Davis_NM:2018}.
Especially, the estimated ratio of the quasiparticle spectral weights between the $yz$ and $xz$ orbitals is very large: $Z_{yz}/Z_{xz}\sim 4$.
Given the
 small splitting between these two orbitals~\citep{Watson_PRB:2016,Yi_PRX:2019}, such a strong orbital selectivity is surprising~\citep{Bascones_PRB:2017}.

To resolve this puzzle, we examine the electron correlation effects in a multiorbital Hubbard model for the nematic phase of FeSe using the $U(1)$ slave-spin theory.
To generate a momentum dependent orbital splitting, besides the momentum independent ferro-orbital order ($\delta_f$), we have also taken into account a $d$- and an $s$-wave bond nematic order ($\delta_d$ and $\delta_s$), which corresponds to nearest-neighboring hopping anisotropy~\citep{Su_JPCM:2015}. We add the Hamiltonian $H_{\rm{nem}}$ that describes the effects of these nematic orders into Eq.~\eqref{Eq:Ham_tot}.
In the momentum space
\begin{eqnarray}\label{Eq:Ham_nem}
 && H_{\rm{nem}} = \sum_k \left[ -2\delta_d (\cos k_x-\cos k_y)(n_{k1}+n_{k2}) \right.\nonumber\\
 && - \left.  2\delta_s (\cos k_x + \cos k_y) (n_{k1}-n_{k2}) + \delta_f (n_{k1}-n_{k2}) \right].
\end{eqnarray}

\begin{figure}[h!]
\begin{center}
\includegraphics[width=12cm]{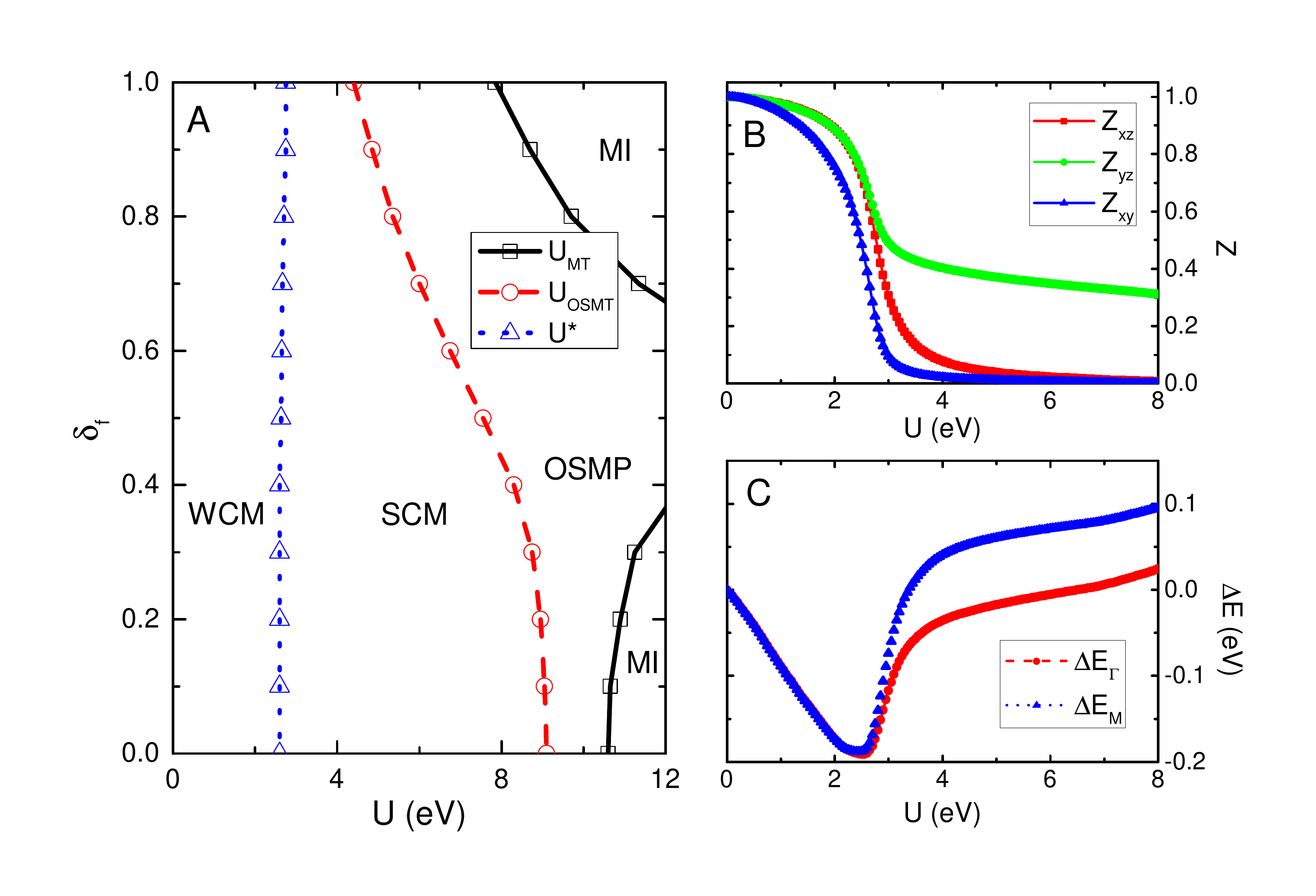}
\end{center}
\caption{(A): Ground-state phase diagram of the five-orbital Hubbard model for FeSe
with a ferro-orbital order $\delta_f$ at $J_{\rm{H}}/U=0.25$. (B): The orbital-selective quasiparticle spectral weights
in the nematic phase with a combined nematic order $\delta_f/4= \delta_d =\delta_s =0.2$ eV and with $J_{\rm{H}}/U=0.25$.
(C): Band splittings at $\Gamma$ ($\Delta E_\Gamma$) and M ($\Delta E_{\rm{M}}$) points of the 2-Fe BZ
in the nematic phase with the same set of parameters as in (B). Adapted from Ref.~\citep{Yu_PRL:2018}.}\label{fig:4}
\end{figure}

By solving the saddle-point equations, we show that the OSMP is promoted by any of these nematic
orders, as illustrated in Fig.~\ref{fig:4}(A). This effect is delicate, because we also find that
the full Mott localization of the system depends on the type and strength of the nematic order~\citep{Yu_PRL:2018}.
Remarkably, we find that, by taking a proper combination of the three
types of nematic order,
the system
 can exhibit a strong orbital selectivity with $Z_{yz}/Z_{xz}\sim 4$
 while, at the same time,
  shows a small band splittings at the $\Gamma$ and M points of the BZ (with $\Delta E_\Gamma$,
  $\Delta E_{\rm{M}}\lesssim 50$ meV) as a result of  a cancelation effect (see Fig.~\ref{fig:4}(B) and (C))~\citep{Yu_PRL:2018}.
These results reconcile the seemingly contradictory ARPES and STM results. The explanation to the unusually large orbital selectivity
 in the nematic phase of FeSe~\citep{Davis_Sci:2017,Davis_NM:2018} further shed light to the understanding
of  the superconductivity in this compound, which we will discussed in
Sec.~\ref{Sec:SCPairing_NemPairing}.

\section{Orbital-selective superconducting pairing}\label{Sec:SCPairing}

In Sec.~\ref{Sec:NormalState} we have discussed the orbital-selective electron correlations in the normal state of FeSCs.
It is natural to ask whether the strong orbital selectivity can affect the pairing symmetry and amplitudes in the superconducting states.
The effects of orbital selectivity on superconductivity are two-fold.
 The orbital selectivity
 modifies the bandstructure from  its noninteracting counterpart.
  This has been verified by ARPES mesurements~\citep{Yi_NC:2015,Liu_PRB:2015,Miao_PRB:2016}.
 In addition, the orbital selectivity influences the
 effective interactions
 projected to the pairing channel.
 In the following, we study these effects
 in a frustrated multiorbital $t$-$J$ model. We show
 that
 any
 interorbital pairing
 has a
 negligible
 amplitude; the structure of the pairing state is then reflected in
 the pairing
 amplitude
 being orbital dependent,
 which is denoted as orbital-selective pairing. In FeSCs, this may give rise to
 superconducting gaps
 with unexpectedly strong anisotropy
 as well as new type of pairing states that
 has no single-orbital counterpart;
 we will discuss how both types of effects play  an important role in
 several iron pnictide and iron chalcogenide compounds~\citep{Yu_PRB:2014,Nica_QM:2017,Hu_PRB:2018}.

\subsection{Superconducting pairing in the multiorbital $t$-$J$ model}\label{Sec:SCPairing_tJ}

The bad metal behavior in the normal state implies strong electron correlations in FeSCs. In strongly correlated systems, the effective superconducting pairing has to avoid the penalty
from
the Coulomb repulsion.
Even though
the parent compound is not a MI,
the superconducting phase
in most cases
is in proximity to an AFM phase. This
suggests
 that the AFM exchange interaction plays a very important role for superconductivity.
It has been shown theoretically that the AFM exchange interaction is enhanced in
the bad metal
($0<Z\ll1$)
 regimes near the Mott transition~\citep{Ding_PRB:2019}.
 In a multiorbital correlated system, similar enhancement is anticipated when the system is close to an OSMP.
 We therefore proceed to
 study the superconducting pairing within the framework of a multiorbital $t$-$J$ model,
 such that the orbital selectivity in the normal state is taken into account.

The effective Hamiltonian of the model
has the following form~\citep{Hu_PRB:2018}
\begin{eqnarray}\label{HamtJ}
H_{eff} &=& \sum_{ij,\alpha\beta\sigma}(\sqrt{Z_{\alpha}Z_{\beta}} t_{ij}^{\alpha\beta}-\tilde{\lambda}_{\alpha}\delta_{\alpha\beta})
f^{\dagger}_{i,\alpha,\sigma} f_{j,\beta,\sigma}
-\sum_{ij,\alpha\beta} J_{ij}^{\alpha\beta}
f_{j,\beta,\downarrow}^{\dagger} f_{i,\alpha,\uparrow}^{\dagger} f_{i,\alpha,\downarrow} f_{j,\beta,\uparrow} \, .
\end{eqnarray}
Here, the bands are renormalized by the quasiparticle spectral weights $Z_{\alpha}$'s;
as we described in the previous section,
the orbital ($\alpha$) dependence of this weight reflects the orbital selectivity in the normal state.
In addition,
 $\tilde{\lambda}_\alpha$ is the effective energy level
 that takes
 into account the correlation effect (see Eq.~\eqref{Eq:Hfmf});
 $J_{ij}^{\alpha\beta}$ refers to the orbital dependent AFM exchange couplings,
 which can be obtained by integrating out the incoherent single-electron excitations via either the slave-rotor~\citep{Ding_PRB:2019} or slave-spin~\citep{Ding_arXiv:2019} approach.
 It is generically a
  matrix in the orbital space.
  However, the interorbital interactions have been found to generate negligible
  interorbital pairing~\citep{Nica_QM:2017}.
  Thus,
  the most important
 terms to the pairing are the
 the diagonal interactions
 in the $t_{2g}$ ($d_{xz}, d_{yz}, d_{xy}$) orbital subspace.
We focus on these interactions
 up to the next nearest
 neighbors.
 We further introduce two ratios to quantify the effects of magnetic frustration and orbital selectivity,
 respectively: $r_L=J_1/J_2$, and $r_O=J^{xy}/J^{xz/yz}$. In principle $r_L$ and $r_O$ can be determined
 from
 the procedure of integrating out high-energy incoherent states~\cite{Ding_PRB:2019}.
 In practice we take them as
  model parameters,
 so that
   a comprehensive understanding on the pairing states
   can be gained.

To study the
superconducting pairing,
we decompose the interaction term in Eq.~\eqref{HamtJ} in the pairing channel by introducing the
pairing fields
 in the real space: $\Delta^{\alpha}_{\textbf{e}}=\frac{1}{\mathcal{N}}\sum_i f_{i,\alpha,\uparrow} f_{i+\textbf{e},\alpha,\downarrow}$,
 where $\textbf{e}\in\{e_x,e_y,e_{x+y},e_{x-y}\}$
refers to a unit vector connecting nearest and next nearest neighboring sites. Transforming to the momentum space we obtain different pairing channels, each of which corresponds to linear combinations of several
pairing
 fields in the real space. In general, the pairing function $\Delta_{g,\alpha,\mathbf{k}}=\Delta_{g,\alpha(\tau_i)} g(\mathbf{k}) (\tau_i)_\alpha$, where $\Delta_{g,\alpha(\tau_i)}$
 is the pairing strength of a particular pairing channel. For the non-degenerate $d_{xy}$ orbital,
 the symmetry of the pairing state is fully determined by the form factor $g(\mathbf{k})$, and the four channels
 are usually denoted as $s_{x^2+y^2}=\cos k_x + \cos k_y$, $d_{x^2-y^2}=\cos k_x - \cos k_y$, $s_{x^2y^2}=\cos k_x \cos k_y$,
 and $d_{xy}=\sin k_x \sin k_y$, respectively. In the tetragonal phase, the degeneracy of $d_{xz}$ and $d_{yz}$ orbitals
 introduces additional pairing channels,  and it is necessary to  use
 the $2\times 2$ Pauli matrices $\tau_i$ in the orbital isospin space to construct the various possibilities.
The pairing channels can be classified according to the one-dimensional irreducible representations of the tetragonal $D_{4h}$
point group to be $A_{1g}$, $B_{1g}$, $A_{2g}$, and $B_{2g}$.
Note that different channels may exhibit the same symmetry,
by combining the structure in both the form factor and the orbital structure.
For example, both the $d_{x^2-y^2}$ wave channel in the $d_{xy}$ orbital and the $s_{x^2y^2} \tau_3$
channel in the $d_{xz/yz}$ orbitals have the $B_{1g}$ symmetry.

\subsection{Orbital-selective pairing in FeSCs}\label{Sec:SCPairing_OSPairing}

Since the discovery of FeSCs, the pairing symmetry of the superconducting state has been one of the most
important questions. The $s$-wave $A_{1g}$ channel has played a particularly important role.
In addition,
 various subleading
 --in some cases, nearly-degenerate--
 pairing channels with compatible symmetry can coexist.

 The notion of orbital-selective pairing
 was introduced~\cite{Yu_PRB:2014}
 in the multiorbital $t$-$J$ model for electron doped NaFeAs.
 With the intra-orbital pairing amplitudes being dominant, this leads to a
 multigap structure, with
 different pairing
 components coming from different orbitals.
Because the orbital composition varies along each pocket as shown in Fig.~\ref{fig:1}(B),
this orbital-selective pairing can give rise to an anisotropic superconducting gap.

\begin{figure}[tbh!]
\begin{center}
\includegraphics[width=12cm]{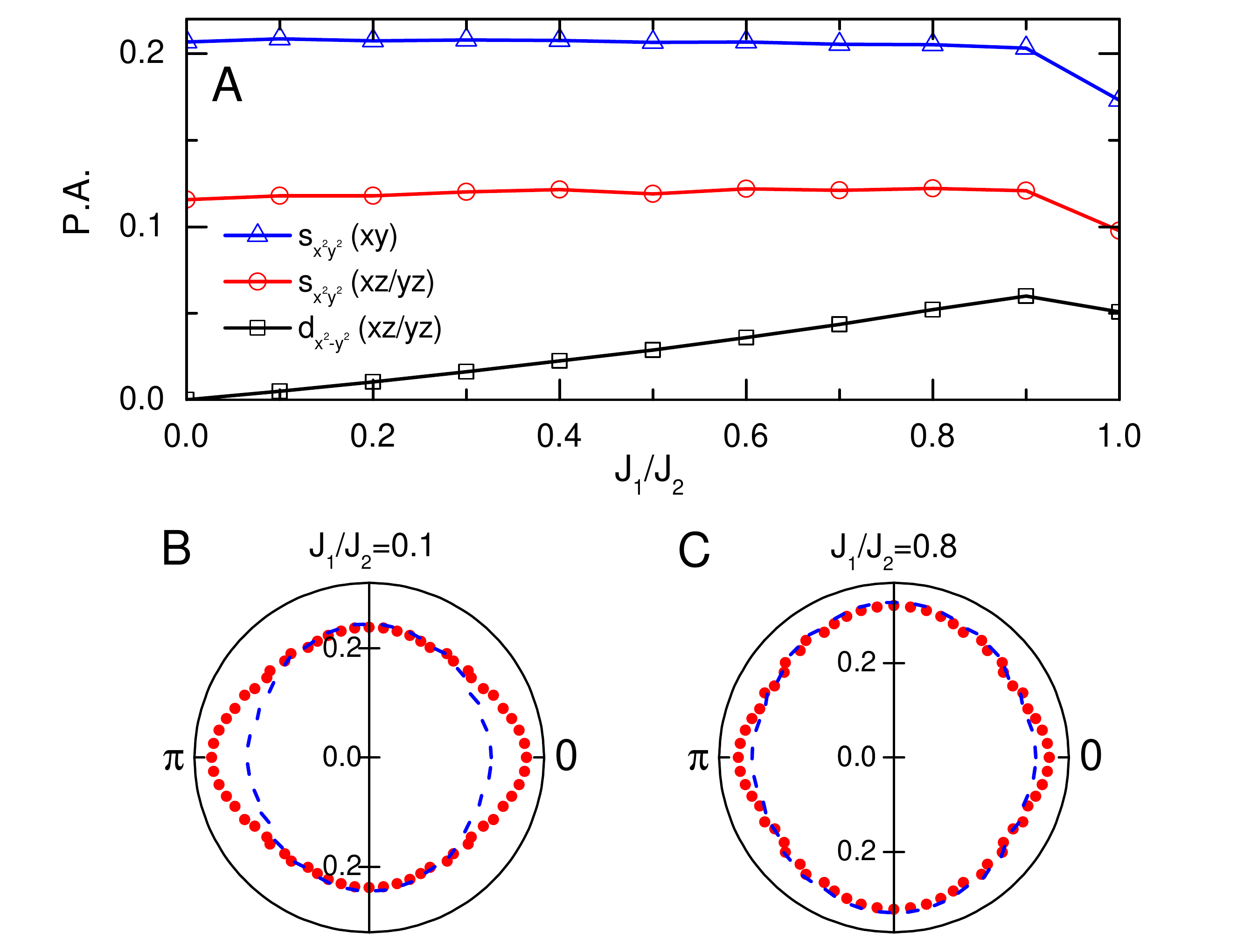}
\end{center}
\caption{(A): Evolution of the leading pairing channels with $J_1/J_2$ in the multiorbital $t$-$J$ model for electron doped NaFeAs. All the channels have the $A_{1g}$ symmetry. P.A. denotes pairing amplitude. (B), (C): Angular dependence of the superconducting gaps (red circles) along the electron pocket centered at $(\pi,0)$ in the same model
at $J_1/J_2 = 0.1$ [in (B)] and $J_1/J_2 = 0.8$ [in (C)], respectively. The blue dashed line is a fit to the single parameter gap function $\Delta_0 \cos k_x \cos k_y$. The deviation from this fit implies a multigap structure arising from the orbital-selective pairing. Adapted from Ref.~\citep{Yu_PRB:2014}.}\label{fig:5}
\end{figure}

For simplicity, the exchange couplings have been assumed to be orbital independent in the calculation, and the pairing state has an $A_{1g}$ symmetry with a full gap along the Fermi surfaces.
The different bandwidths and electron fillings of the $d_{xy}$ and $d_{xz/yz}$ orbitals still make the pairing to be
orbital selective. As shown in Fig.~\ref{fig:5}(A), the pairing amplitude of the leading channel, the $s_{x^2y^2}$ channel in the $d_{xy}$ orbital, is much larger than those of the subleading channels in the $d_{xz/yz}$ orbitals.
This orbital-selective pairing is reflected in
an anisotropy of the superconducting gap along the electron pocket, as shown in Fig.~\ref{fig:5}(B).
It turns out that
 the superconducting gap strongly depends on the magnetic frustration.
 For $J_1/J_2=0.8$ as illustrated in Fig.~\ref{fig:5}(C), the gap becomes almost isotropic.
 Here, several competing channels become active, which makes
 the overall contributions to the gap from the $d_{xz/yz}$ and $d_{xy}$ orbitals comparable.
 In experiments, an almost isotropic superconducting gap has been reported for the optimally electron doped NaFeAs,
 but the gap becomes strongly anisotropic along the electron pocket in the underdoped compound~\citep{Ge_PRX:2013}.
 This behavior
  is understood by the strong orbital-selective pairing~\citep{Yu_PRB:2014}, which also
 splits into two the
 neutron resonance peak
 as a function of energy
 in the superconducting state~\citep{Zhang_PRL:2013,Zhang_PRB:2015}.
 The orbital-selective pairing is also being explored in other FeSCs~\citep{Tian_Dai_PRB:2020}.

\begin{figure}[tbh!]
\begin{center}
\includegraphics[width=17cm]{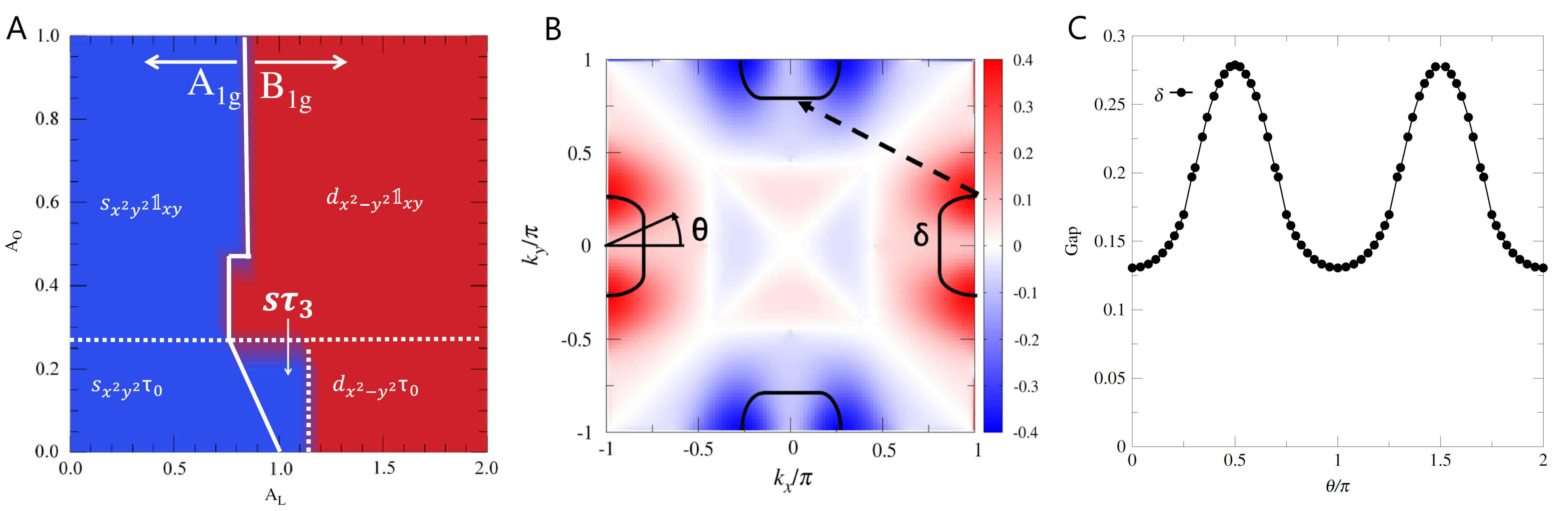}
\end{center}
\caption{(A): Pairing phase diagram of the multiorbital $t$-$J$ model for alkaline iron selenides. The blue shaded area
corresponds
 to dominant pairing channels with an $s_{x^2y^2}$ form factor while the red shading covers those with a $d_{x^2-y^2}$ form factor. The continuous line separates regions where the pairing belongs to the $A_{1g}$ and the $B_{1g}$ representations respectively. The orbital-selective $s\tau_3$ pairing occurs for $A_O < 1$, $A_L$ near $1$. (B): The Fermi surface (solid line) and the sign-change intra-band pairing function of the corresponding band for the alkaline iron selenides. The dashed arrow indicates the $\mathbf{q}=(\pi,\pi/2)$ wave-vector associated with the spin resonance found in experiment. (C): Superconducting gap along the electron pocket. In (B) and (C) the dominant $s\tau_3$ pairing state is stabilized with parameters $J_2=1.5$, $A_O = 0.3$, and $A_L = 0.9$. Adapted from Ref.~\citep{Nica_QM:2017}.}\label{fig:6}
\end{figure}

Besides the gap anisotropy, strong orbital selectivity may give rise to novel pairing states. In a multiorbital $t$-$J$ model with orbital independent exchange couplings, the dominant pairing symmetry has been found to be either an $s$-wave $A_{1g}$ channel when $J_2$ is dominant, or a $d$-wave $B_{1g}$ channel for dominant $J_1$ coupling~\cite{Goswami_EPL:2010,Yu_NC:2013}.
In the regime where the two types of pairing states are quasi-degenerate,
a novel orbital-selective pairing state, $s\tau_3$ pairing, with $s$-wave form factor but $B_{1g}$
symmetry in the $d_{xz/yz}$ orbital subspace can be stabilized as the leading pairing channel~\citep{Nica_QM:2017}
 (see Fig.~\ref{fig:6}(A)).
However, its nontrivial orbital structure
makes it distinct from
the other two pairing channels, and this successfully explains the unconventional superconductivity in alkaline iron chalcogenides~\citep{Nica_QM:2017}: It produces a full gap but the pairing function has a sign change
between the two electron pockets, which causes a spin resonance around the wave vector $(\pi,\pi/2)$, as shown
in Fig.~\ref{fig:6}(B) and (C). These features are compatible with both the ARPES and neutron scattering   measurements~\citep{Mou_PRL:2011,Wang_EPL:2011,Wang_EPL:2012,Park_PRL:2011,Friemel_PRB:2012}.

Importantly, the $s\tau_3$ pairing corresponds to an irreducible representation of the crystalline point group. Nonetheless,
in the band basis, this  multiorbital
superconducting state
has the form of a mutliband $d+d$ pairing. The latter allows the state to contrast with the more familiar $d+id$ pairing
in a way that the $^3$He-B superfluid state contrasts with $^3$He-A state~\citep{Nica_arXiv:2019}.
This understanding elevates the $d+d$ pairing to the status of a natural multiorbital pairing state.
We note in passing
 that an analogous $s\tau_3$ pairing has been constructed
 for the first unconventional superconductor CeCu$_2$Si$_2$ \citep{Nica_arXiv:2019},
 which provides
 a natural understanding of the recently discovered
 low-temperature behavior in this heavy fermion
 compound~\citep{Pang_PNAS:2018}.

\subsection{Orbital-selective pairing in the nematic phase of iron selenide}\label{Sec:SCPairing_NemPairing}

As we discussed in Sec.~\ref{Sec:NormalState_NemOSM}, recent STM measurements in the nematic phase
of FeSe have uncovered not only a surprisingly large difference between the quasiparticle weights of the $d_{xz}$
and $d_{yz}$ orbitals, but also
an
unusually strong anisotropy of the superconducting gap~\cite{Davis_Sci:2017,Davis_NM:2018}.
 These experimental findings
 provide evidence for
 a
strongly orbital-selective superconducting state.

\begin{figure}[tbh!]
\begin{center}
\includegraphics[width=10cm]{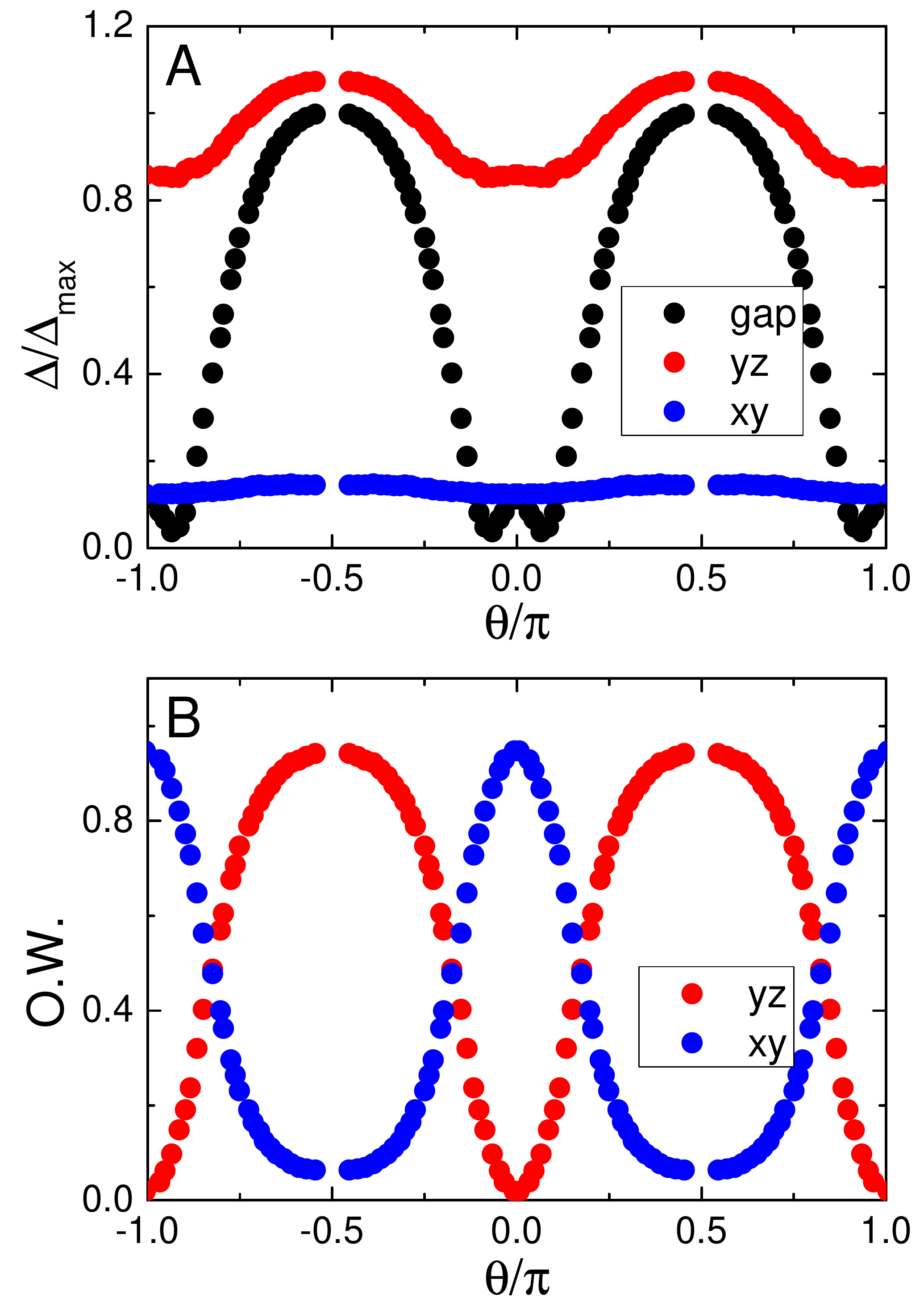}
\end{center}
\caption{(A): Overall (blue symbols) and orbital resolved superconducting
gaps along the M$_x$ electron pocket. (B): Weight
distributions of the $d_{xy}$ and $d_{yz}$ orbitals along the M$_x$ electron pocket. Adapted from Ref.~\citep{Hu_PRB:2018}}\label{fig:7}
\end{figure}

Theoretically, the pairing structure in the nematic phase of FeSe
has
been investigated within the framework of the multiorbital $t$-$J$ model in Eq.~\eqref{HamtJ}. The slave-spin claculation~\cite{Yu_PRL:2018} produces $Z_{xz}:Z_{yz}:Z_{xy} = 1 : 4 : 0.5$,
an orbital-selective pairing  was shown, with the leading pairing channel in the $d_{yz}$ orbital.
Taking into account the mixed orbital character of both hole and electron pockets,
such an orbital-selective pairing naturally leads to a large gap anisotropy as shown in Fig.~\ref{fig:7}.
This orbital-selective pairing
not only provides
the understanding of the experimental observations~\citep{Davis_Sci:2017,Liu_PRX:2018,Rhodes_Kim_FeSe_PRB:2018,Kushnirenko_Borisenko_PRB:2018},
but also sheds new light on the interplay among the pairing state, Mott physics, and the nematic order,
 all of which appear to be important ingredients for the unconventional superconductivity in FeSCs.
 Experimentally, other signatures of orbital-selective superconductivity in nematic FeSC are also being explored
~ \citep{Chen_Dai_NatMater:2019}.

\section{Summary and outlook}\label{Sec:Summary}

Since the discovery of superconductivity in FeSCs,
clarifying the underlying microscopic
 physics of these materials has been the goal of extensive research, and
 considerable progress has
 been achieved.
 By now it has become abundantly clear that
   electron correlations play a key role.
  This includes both the Hubbard and Hund's couplings,
  which
  combine to cause
  the normal state of the
  FeSCs to be a bad metal in
  proximity to a Mott transition.
  Theoretical studies on
  the pertinent
  microscopic models for the FeSCs not only confirms the existence of the bad metal in the phase diagram,
  but also reveals
  a strong orbital selectivity
   in this phase, which is anchored by an orbital-selective Mott phase.
    In this manuscript we have reviewed recent theoretical progress on the orbital selectivity.
    It has been found that the orbital selectivity
    not only is a universal property of the normal state of FeSCs, but also shows
    intriguing  interplay with the nematicity.
  Equally important,   it can strongly affect the superconducting states of the system.

It is worth reiterating
 that the FeSCs consist of a large family of materials, and superconductivity has been found over a broad range of tuning parameters,
 such as pressure and electron filling. For example,
 many electron-doped iron chalcogenides have
 a simpler Fermi surface, with
 only electron pockets, and the electron filling is $n\sim6.1-6.2$.
 Even so, superconductivity is also discovered and, indeed, it is in this category of materials that the highest-$T_c$
 FeSC belongs. Here,
  the strong orbital selectivity is shown to be universal in these systems, and has been extensively studied.

\begin{figure}[tbh!]
\begin{center}
\includegraphics[width=12cm]{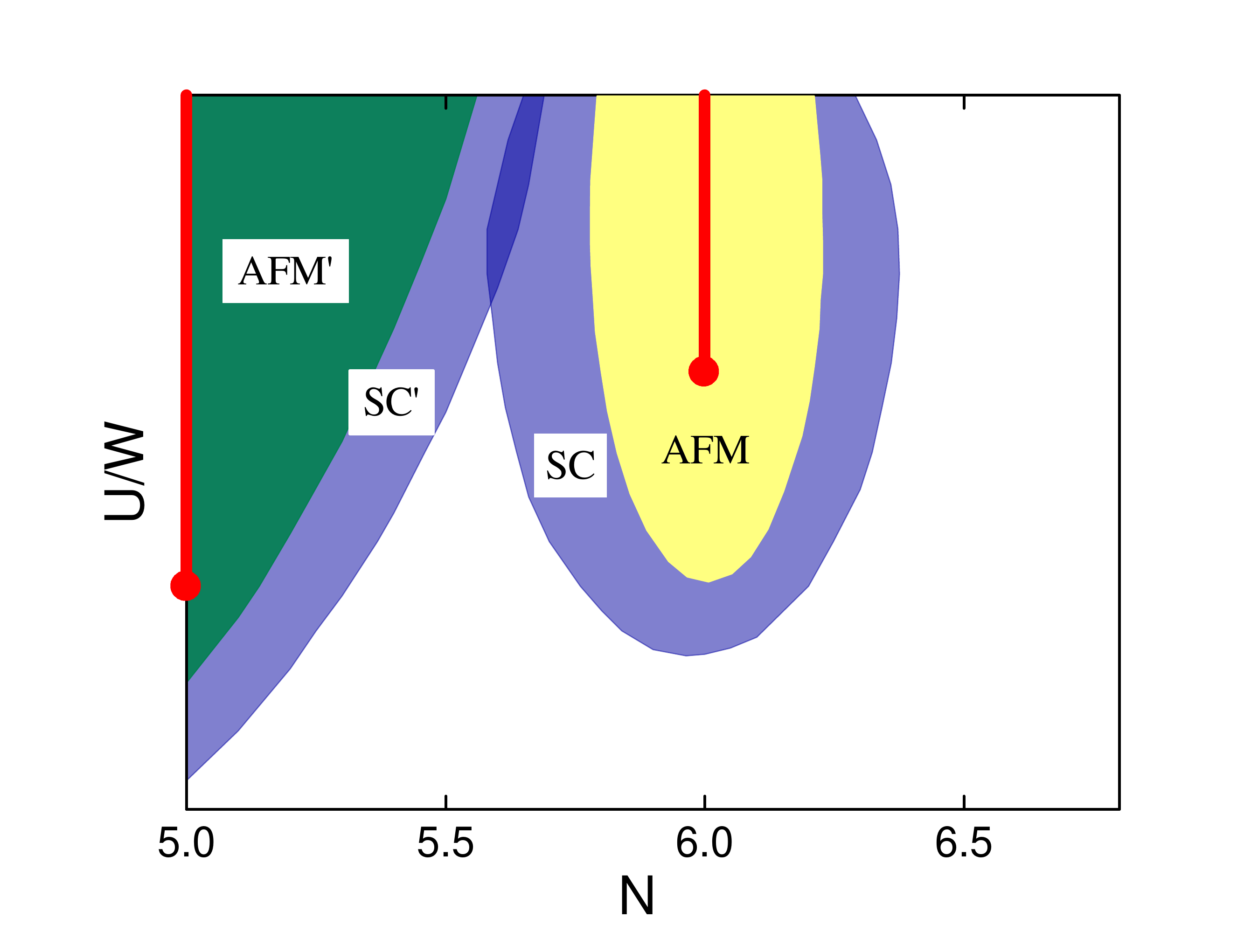}
\end{center}
\caption{Schematic phase diagram as a function of the $U/W$ (the ratio of the Coulomb interaction to bandwidth).
Here the red lines denote the MIs at electron filling $n=6$ and $n=5$, respectively. AFM marks the $(\pi,0)$ AFM order
near the $n=6$ MI, while AFM$^\prime$ represents the AFM order near the $n\rightarrow 5$
limit~\citep{Yu_COSSMS:2013}.
SC and SC$^\prime$ denote two superconducting states near the two AFM phases. Adapted from Ref.~\citep{Wang_PRB:2019}.}\label{fig:8}
\end{figure}

 Also of note is the case of extremely hole-doped iron pnictides, which likewise displays superconductivity.
 A prototype class of materials in this category is
 $A$Fe$_2$As$_2$ ($A$=K, Rb, Cs), which
 contains hole pockets only, and the electron filling is at $n=5.5$.
 As illustrated in Fig.~\ref{fig:8}, they are far from both the $n=6$ and $n=5$ MIs.
 Although the superconducting $T_c$ observed in these systems is much lower than that of the
 other iron-based systems, a number of
 recent experiments prompt the consideration of
a completely different type of antiferromagnetism and nematicity~\citep{Feng_NC:2019,Shibauchi_arXiv:2018,Wang_PRB:2016,WuChen_arXiv:2016,Horigane_SR:2016,Eilers_PRL:2016,
Wiecki2020},
which are
possibly associated with the Mott physics in the $n=5$ limit.
In particular, the $d_{xy}$ orbital is shown to
be
closer to the Mott localization
with reducing the total electron filling. Therefore, a systematic study on the evolution of orbital-selective correlations with doping from $n=6$ to $n=5$ would be important in clarifying the underlying physics and the connection to superconductivity in these heavily hole doped materials.

A topic of considerable
recent interest in the area of
FeSCs is the indication for a topologically nontrivial bandstructure and the possible Majorana zero mode in
the superconducting
iron chalcogenides~\citep{Zhang_Sci:2018}. This highlights the important role of spin-orbit coupling in these systems.
Given the compelling evidence for the strongly orbital selective correlations we have discussed here,
it would be highly desirable to clarify
how the interplay between the correlation effects and the spin-orbit coupling
would affect the topological properties of the electronic bandstructure.
Such efforts promise to elucidate the extent to which the topological bandstructure develops in the various families of FeSCs.

\section*{Author Contributions}

All authors contributed to the writing of the manuscript.

\section*{Funding}
Work at Renmin University has been supported by the National Science Foundation of China
Grant number 11674392, the Ministry of Science and Technology of China, National Program on Key Research Project Grant number 2016YFA0300504, and the Fundamental Research Funds for the
Central Universities and the Research Funds of Remnin University
of China Grant number 18XNLG24.
Work at Rice has been supported by the DOE BES Award \#DE-SC0018197 and the Robert A. Welch Foundation Grant No. C-1411.
Work at Los alamos was
 carried out under the auspices of the U.S. Department of Energy (DOE)  National Nuclear Security Administration under Contract No. 89233218CNA000001, 
and was supported by the LANL LDRD Program. 
Q.S. acknowledges the support of the NSF Grant No.\,PHY-1607611
at the Aspen Center for Physics.

\section*{Acknowledgments}
We thank the late E. Abrahams, R. J. Birgeneau, P. C. Dai, W. Ding, P. Goswami, K. Jin, C. L. Liu, D. H. Lu, X. Y. Lu, P. Nikolic,
Z.-X. Shen,  Y. Song, M. Yi, and W. Yu for useful discussions.
R.Y. acknowledges the hospitality of the T.D. Lee Institute.





\section*{Figures}

\end{document}